\documentclass[a4paper,11pt]{article}
\ifdefined\pdfoutput\pdfoutput=1\fi 

\usepackage{jcappub} 

\usepackage[T1]{fontenc} 
\usepackage{xcolor}

\usepackage{graphicx}
\makeatletter
\def\caption@documentclass{standard}
\makeatother
\usepackage{subcaption}
\usepackage{float}
\usepackage{array}
\usepackage[normalem]{ulem}

\title{Disc-Based Estimation of the Local Dark Matter Density and Velocity Distribution in IllustrisTNG50}

\author[a]{Hatsume~Chujo,}
\author[b,c]{Shogo~Masaki,}
\author[a]{Keiko~I.~Nagao}
\author[a,d]{and~Takuho~Nakabayashi}

\affiliation[a]{Okayama University of Science, 1-1 Ridaicho Kita-ku, Okayama, Okayama 700-0005
Japan}
\affiliation[b]{Department of Information Engineering and Institute for Advanced Studies in Artificial Intelligence, Chukyo University, 101 Tokodachi, Kaizu-cho, Toyota, Aichi, 470-0393 Japan}
\affiliation[c]{Department of Physics, Nagoya University, Furo-cho, Nagoya, Aichi, 464-8602 Japan}
\affiliation[d]{SOKENDAI, 1-1 Oho, Tsukuba, Ibaraki 305-0801, Japan}

\emailAdd{r25nmc6vn@ous.jp}
\emailAdd{shogo.masaki@gmail.com}
\emailAdd{nagao@ous.ac.jp}
\emailAdd{takuho@post.kek.jp}

\abstract{
We study the dark matter (DM) density and velocity distribution in the solar neighborhood of Milky Way-like galaxies using the TNG50-1 run of the IllustrisTNG simulation suite. Milky Way-like galaxies are selected according to both their halo mass and their bulge-to-total stellar mass ratio, and the solar-neighborhood region within each galaxy is identified from the positions and number density of stars, so that DM particles associated with the galactic disc are extracted while contributions from the halo component are suppressed. Averaging over the selected galaxies, we obtain a local DM density of $\rho \simeq 0.47\pm0.02~{\rm GeV\,cm^{-3}}$, consistent with the range inferred from observations. The disc-based extraction gives a higher peak density, $0.41\pm0.02~{\rm GeV\,cm^{-3}}$, than the value $0.34\pm0.02~{\rm GeV\,cm^{-3}}$ from a simple distance-based extraction, probably because it excludes the low-density DM in the halo. The width of the density distribution, on the other hand, is nearly unchanged. The velocity distribution is well described by an isotropic Maxwellian with a peak velocity of about $226~{\rm km~s^{-1}}$, close to the value assumed in the standard halo model. The differences in the density and velocity distributions between our disc-based and simple distance-based extractions are not so large as to significantly affect the interpretation of direct-detection experiments.}

\begin{document}
\maketitle
\raggedbottom

\section{Introduction}
Dark matter (DM) has been supported by a wide range of evidence, from galactic rotation curves to gravitational lensing effects, large-scale structure of the universe, and the cosmic microwave background (CMB), making its existence unquestionable. Many proposals have been made for DM candidates, including particle physics candidates represented by Weakly Interacting Massive Particles (WIMPs) and axion-like particles (ALPs), as well as astronomical candidates such as Primordial Black Holes (PBHs) \cite{Bozorgnia:2024pwk}. For experimental and observational tests of these DM candidates, including direct searches for WIMPs, the mass density and velocity distribution of DM at the Earth are crucial input quantities that can significantly affect the interpretation of the results. In particular, it has been pointed out that, in direct detection experiments for WIMPs, the resulting constraints on the particle physics parameter space can vary significantly depending on these parameters \cite{Ling:2009eh, Bozorgnia:2016ogo,Kelso:2016qqj,Sloane:2016kyi,Bozorgnia:2017brl,Vogelsberger:2008qb, Folsom:2025lly}. The local DM density is evaluated to be approximately $\rho\simeq 0.2\text{--}0.6$ ${\rm GeV}\,{\rm cm}^{-3}$ based on 
observations of stars' motion in the solar neighborhood and galactic rotation curve observations \cite{Read:2014qva, deSalas:2020hbh}.
Evaluations from several cosmological simulations have yielded comparable values of $\rho\simeq 0.3\text{--}0.7$ ${\rm GeV}\,{\rm cm}^{-3}$ \cite{Bozorgnia:2016ogo, Sloane:2016kyi, Kelso:2016qqj, Staudt:2024tdq, Folsom:2025lly}.
On the other hand, regarding the velocity distribution of DM, although many cosmological simulations show results that are Maxwellian-like, some studies have reported deviations from the Maxwellian distribution and anisotropies caused by residual debris from galaxy mergers or co-rotation effects with the stellar disc \cite{Bozorgnia:2016ogo, Kelso:2016qqj, Necib:2018igl, Santos-Santos:2023ubx, Staudt:2024tdq, Folsom:2025lly, Shpigel:2025ulk, Zhang:2026qnl}.

In previous simulation studies discussing the mass density and velocity distribution of DM in the solar neighborhood, various criteria have been tested to select the solar neighborhood region. Among these, some studies have conducted their analyses under less stringent selection criteria, either for choosing Milky Way (MW)-like galaxies or for selecting solar-neighborhood-like regions within each galaxy.
For instance, while certain studies impose only a mass condition requiring a halo mass close to that of the MW \cite{Vogelsberger:2008qb, Sloane:2016kyi}, others additionally impose conditions on the galactic rotation curve, the stellar mass, or galactic environment \cite{Bozorgnia:2016ogo, Folsom:2025lly}. Similarly, for the selection of a solar-neighborhood-like region within a galaxy, some studies require only a galactocentric distance of about 8.3 kpc \cite{Vogelsberger:2008qb, Zemp:2008gw}, while others additionally restrict the region to the disc to avoid contamination from halo DM \cite{Pato:2010yq, Staudt:2024tdq}. An inappropriate selection may yield a region that does not reflect the true conditions of the solar neighborhood, leading to an unreliable estimate of the local DM properties.

Motivated by these considerations, we evaluate the local DM density and velocity distribution in the solar neighborhood by analyzing DM properties in more realistically selected solar-neighborhood-like regions of simulated galaxies. To do so, we use public data from the IllustrisTNG project \cite{Nelson:2018uso, tng50, tngproject}. It has been reported that IllustrisTNG successfully reproduces many observables, such as the stellar mass function of galaxies, galaxy morphology, galaxy clustering, and the statistical properties of satellite galaxies \cite{Springel:2017tpz}.
To obtain properties of DM in the solar neighborhood from the simulation data, it is necessary to select MW-like galaxies and further choose solar-neighborhood-like regions within those galaxies. In this study, in addition to selecting MW-like galaxies under strict criteria, we employ a disc-based extraction method utilizing stellar positional information and stellar density to identify 
solar-neighborhood-like regions.

This paper is organized as follows. First, in Sec.~\ref{sec:method}, we describe in detail the selection criteria for MW-like galaxies and solar-neighborhood-like regions analyzed in this study. In addition, for comparison, we perform data extraction under conventionally used criteria. In Sec.~\ref{sec:results}, we analyze the mass density and velocity distributions of the local DM obtained under these criteria 
and compare the results across the methods. We conclude the paper in Sec.~\ref{sec:conclusion}.

\section{Simulation Data and Selection Criteria}
\label{sec:method}

\subsection{The simulation}
We use the TNG50-1 run of the IllustrisTNG simulation suite \cite{Nelson:2018uso}, a cosmological magnetohydrodynamical simulation performed with the moving-mesh code AREPO \cite{Springel:2009aa}.
The simulation volume is a cube with a side length of 
$51.7$ comoving-Mpc under periodic boundary conditions, 
containing $2160^3$ DM particles and $2160^3$ initial gas cells. The baryonic mass resolution is $8.5 \times 10^4$ $M_\odot$, the DM particle mass resolution is $4.5 \times 10^5$ $M_\odot$, and the gravitational softening length for stars and DM particles is $288$\,pc at $z=0$. The simulation adopts the {\it Planck} 2015 cosmological parameters, $\Omega_{\rm m} = 0.3089$, $\Omega_\Lambda = 0.6911$, $\Omega_{\rm b} = 0.0486$, and $h = 0.6774$ \cite{Planck:2015fie}. We use the snapshot, group catalog
and supplemental data of galaxy morphologies and bar properties \cite{Zana22} at $z=0$. The snapshot provides not only the positions, velocities, and particle IDs but also the local densities of individual DM particles via \texttt{SubfindDMDensity}. For star particles, we use the 
positions and \texttt{StellarHsml} values, from which the stellar number density is calculated. The group catalogs provide galaxy properties such as the halo mass and stellar mass, while the morphology catalog supplements the bulge and pseudo-bulge mass ratios to the total stellar mass.

In IllustrisTNG, halos are identified as Friends-of-Friends (FoF) groups \cite{Davis:1985rj},
and self-gravitationally bound substructures within them  are identified as subhalos by the SUBFIND algorithm \cite{Springel:2000qu}. We identify the central galaxy of each FoF halo with the primary subhalo indexed
by \texttt{GroupFirstSub}, and regard these central galaxies as candidates for MW-like galaxies. For each candidate, we characterize the mass by $M_{200{\rm c}}$ of its host FoF halo, the total mass enclosed within $R_{200{\rm c}}$, where the mean enclosed density is $200$ times the critical density of the Universe. In the TNG group catalog, this quantity is provided as \texttt{Group\_M\_Crit200}. This definition matches the one adopted for the observational MW mass estimate used in
Sec.~\ref{sec:criteria}.

\subsection{Selection of MW-like Galaxies and Extraction of the Local DM}
\label{sec:criteria}

To identify MW-like galaxies among 
the candidates described in the previous subsection, we consider the following two criteria.
\paragraph*{Criterion I. Halo mass}
\leavevmode\\
The halo mass of the MW within $R_{200{\rm c}}$ has been observationally estimated to be $M_{\rm 200c} \simeq 10^{12}M_\odot$ \cite{McMillan:2016jtx, Callingham:2018vcf, Cautun:2019eaf, Deason:2020kph, Sun:2022qdh, Kravtsov:2024lmc}. Motivated by these estimates, we adopt the broader halo mass range of $M_{\rm 200c} = 0.8 \times 10^{12} M_\odot \text{--} 2.0 \times 10^{12} M_\odot$, which matches the selection adopted in Ref.~\cite{Font:2020}.

\paragraph*{Criterion II. Bulge-to-total stellar mass ratio}
\leavevmode\\
Imposing Criterion I, the mass condition, alone would not constrain the galaxy morphology and could include galaxies whose bulge-to-total stellar mass ratios $B/T$ are substantially larger or smaller than that of the actual MW.
We therefore additionally require $B/T$ to lie in the range $0.112 \leq B/T \leq 0.206$. This range corresponds approximately to twice the quoted lower and upper uncertainties around the observational estimate $B/T=0.150^{+0.028}_{-0.019}$ \cite{Licquia:2014rsa}. 
We define the bulge-to-total stellar mass ratio of the candidate galaxies as
\begin{equation}
  B/T = f_{\rm bulge} + f_{\rm pseudo\mathchar`-bulge},
\end{equation}
where $f_{\rm bulge}$ and $f_{\rm pseudo\mathchar`-bulge}$ are the stellar mass fractions assigned to the bulge and pseudo-bulge components, respectively, in the supplementary kinematic-morphology catalog \cite{Zana22}\footnote{We note that this kinematic definition is not strictly identical to the observational bulge-plus-bar decomposition used to estimate the MW value in Ref.~\cite{Licquia:2014rsa}. We nevertheless adopt it as a practical proxy for the central stellar component in the simulated galaxies.}. In what follows, we refer to the joint application of Criteria I and II as the stringent selection, and to the application of Criterion I alone as the less stringent selection.

\leavevmode\\
Next, within the selected galaxies, we extract the DM in solar-neighborhood-like regions under the following three 
methods summarized in Table~\ref{tab:three_methods_comparison_overview}.

\paragraph*{Method A. Stringent galaxy selection with disc-based DM extraction}
\leavevmode\\
As MW-like galaxies, we adopt those satisfying the stringent selection criteria (the mass and $B/T$ conditions). The DM in the solar-neighborhood-like region is then extracted as follows. The solar system is located at a distance of about $8.3$ kpc from the galactic center \cite{Gillessen:2008qv, McMillan:2011wd}. Extracting based on that distance would yield DM particles located in a spherical shell centered on the galactic center. If the DM density and velocity distribution near the disc plane differ from those away from the disc plane, a spherical-shell selection may not accurately represent the solar neighborhood. For each galaxy, we therefore calculate the stellar number density of all star particles and fit its logarithmic distribution with three Gaussian components, which we associate with the halo, disc, and bulge populations to determine the stellar number density range associated with the disc, as detailed in Appendix~\ref{sec:gaussian}. We then select stars that both fall within the density range and are located at galactocentric distances of $7\text{--}9$ kpc. Finally, we extract the DM particles located within $0.1$ kpc of these selected disc stars and use them in the analysis. With this method, we expect to extract the DM particles located in the vicinity of the stellar disc at about 8.3 kpc from the galactic center in a ring-like shape. Method A imposes both the mass and $B/T$ conditions on the MW-like galaxy selection and additionally accounts for the galactic disc structure in selecting the solar-neighborhood-like region. It is therefore the most stringent of the three methods, and we adopt it as the baseline in this study.

\paragraph*{Method B. Less stringent galaxy selection with disc-based DM extraction}
\leavevmode\\
In this method, the MW-like galaxies are required to satisfy the less stringent selection criteria (the mass condition alone), a less restrictive condition than in Methods A and C. The solar-neighborhood-like region is extracted using exactly the same procedure as in Method A: the stellar number density distribution of all star particles is fitted with three Gaussian components to determine the density range associated with the disc. Stars that simultaneously satisfy both the density range and are located at $7\text{--}9$ kpc from the galactic center are selected, and DM particles within $0.1$ kpc of these stars are extracted.\footnote{For one of the 145 MW-like galaxy candidates, the stellar density distribution could not be separated into the three components corresponding to the bulge, disc, and halo. We therefore excluded this galaxy from the analysis and used the remaining 144 galaxies in Method B.}
By relaxing only the selection criteria for MW-like galaxies while keeping the DM extraction identical to Method A, Method B serves to investigate the effect of the galaxy selection criteria on the evaluation of the local DM density and velocity distribution.

\paragraph*{Method C. Stringent galaxy selection with distance-based DM extraction}
\leavevmode\\
As in Method A, the MW-like galaxies are required to satisfy the stringent selection criteria (the mass and $B/T$ conditions).
For the local region, however, we use neither the stellar position information nor the stellar density; instead, we select DM particles whose distance from the galactic center is $7\text{--}9$ kpc and analyze those particles. Since this method does not account for the galactic disc structure and targets DM in a spherical shell, comparing it with Method A reveals the impact of the selection method for the solar-neighborhood-like region on the evaluation of the local DM density and velocity distribution.

In Methods A and B, the same DM particle may be extracted multiple times from neighboring stars. Duplicate DM particles are identified and removed using their unique \texttt{ParticleIDs}. The local DM density is taken from the \texttt{SubfindDMDensity} of each selected DM particle of the TNG50-1 snapshot and converted to units of ${\rm GeV\,cm^{-3}}$. This quantity is a local SPH estimate of the DM mass density evaluated from neighboring DM particles. The DM velocity distribution is constructed from the magnitudes of the particle velocities relative to the bulk velocity of the host subhalo, which is obtained from \texttt{SubhaloVel} in the group catalog.

\begin{table}[tbp]
  \centering
  \small
  \caption{Summary of the three methods compared in this study. Here, $N_{\rm gal}$ and $N_{\rm DM}$ denote the numbers of selected MW-like galaxies and extracted DM particles, respectively.}

  \label{tab:three_methods_comparison_overview}
  \begin{tabular}{|>{\centering\arraybackslash}m{0.08\linewidth}|m{0.20\linewidth}|m{0.36\linewidth}|>{\centering\arraybackslash}m{0.08\linewidth}|>{\centering\arraybackslash}m{0.09\linewidth}|}
    \hline
    \multicolumn{1}{|>{\centering\arraybackslash}m{0.08\linewidth}|}{\rule{0pt}{2.8ex}Method} & \multicolumn{1}{>{\centering\arraybackslash}m{0.20\linewidth}|}{\shortstack{\rule{0pt}{2.8ex}MW-like\\Galaxy Criteria}} & \multicolumn{1}{>{\centering\arraybackslash}m{0.36\linewidth}|}{\shortstack{\rule{0pt}{2.8ex}Solar-neighborhood-like\\Region Criteria}} & \multicolumn{1}{>{\centering\arraybackslash}m{0.08\linewidth}|}{\rule{0pt}{2.8ex}$N_{\rm gal}$} & \multicolumn{1}{>{\centering\arraybackslash}m{0.09\linewidth}|}{\rule{0pt}{2.8ex}$N_{\rm DM}$} \\
    \hline
    A & \raggedright Stringent (mass\,$+\,B/T$) & DM particles near stars on the disc identified by stellar density & $26$ & $150{,}785$\\
    \hline
    B & \raggedright Less~stringent (mass) & DM particles near stars on the disc identified by stellar density & $144$ & $777{,}039$\\
    \hline
    C & \raggedright Stringent (mass\,$+\,B/T$) & DM particles at $7<r<9$ kpc from the galactic center & 
    $26$ & $792{,}848$\\
    \hline
  \end{tabular}
\end{table}

\subsection{Comparison of Extraction Methods}
\begin{figure}[tbp]
\centering
\begin{subfigure}{0.45\textwidth}
    \centering
    \includegraphics[width=\linewidth]{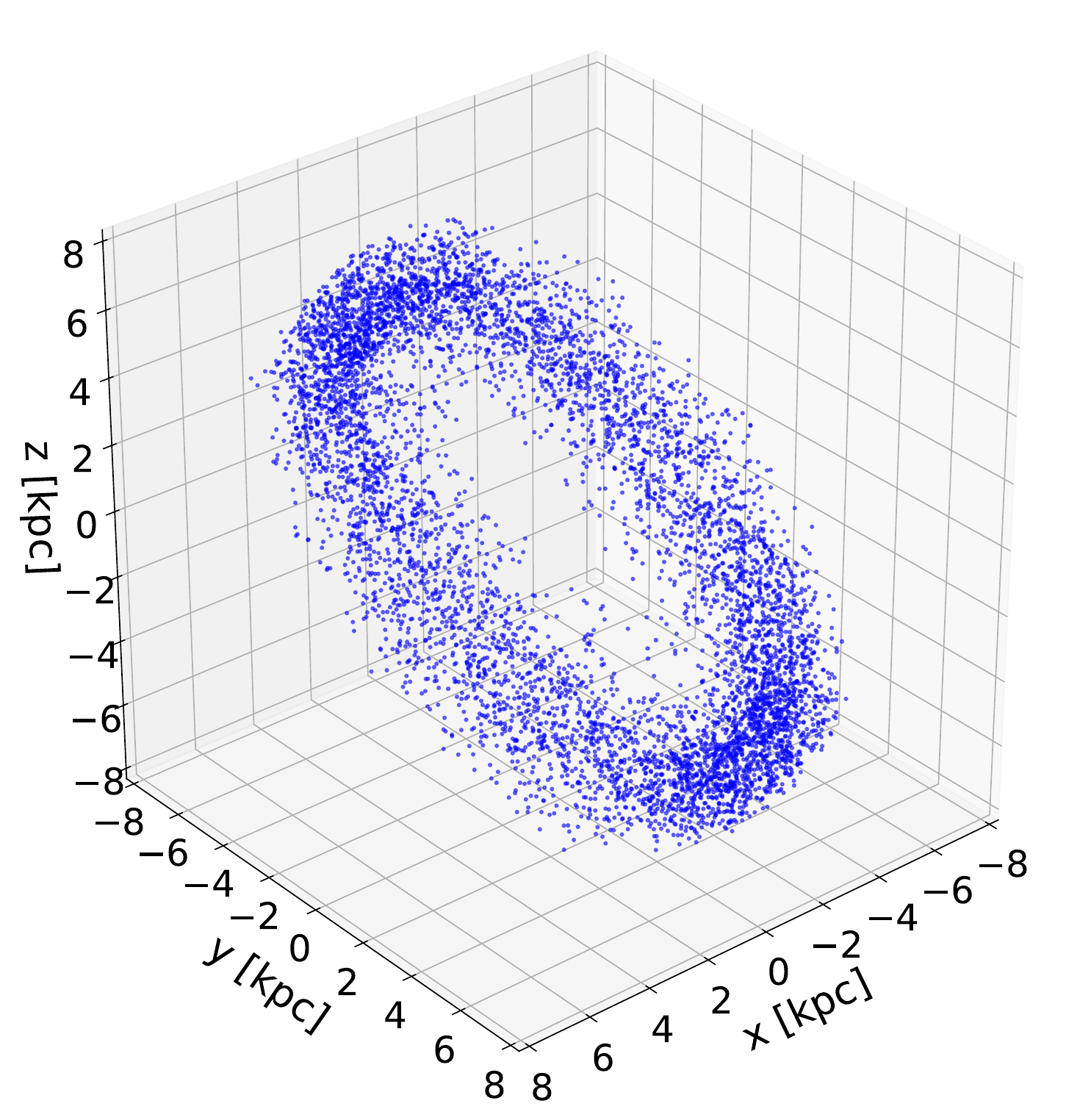}
    \caption{Methods A and B}
    \label{fig:method_compare_3d_AB}
\end{subfigure}
\hfill
\begin{subfigure}{0.45\textwidth}
    \centering
    \includegraphics[width=\linewidth]{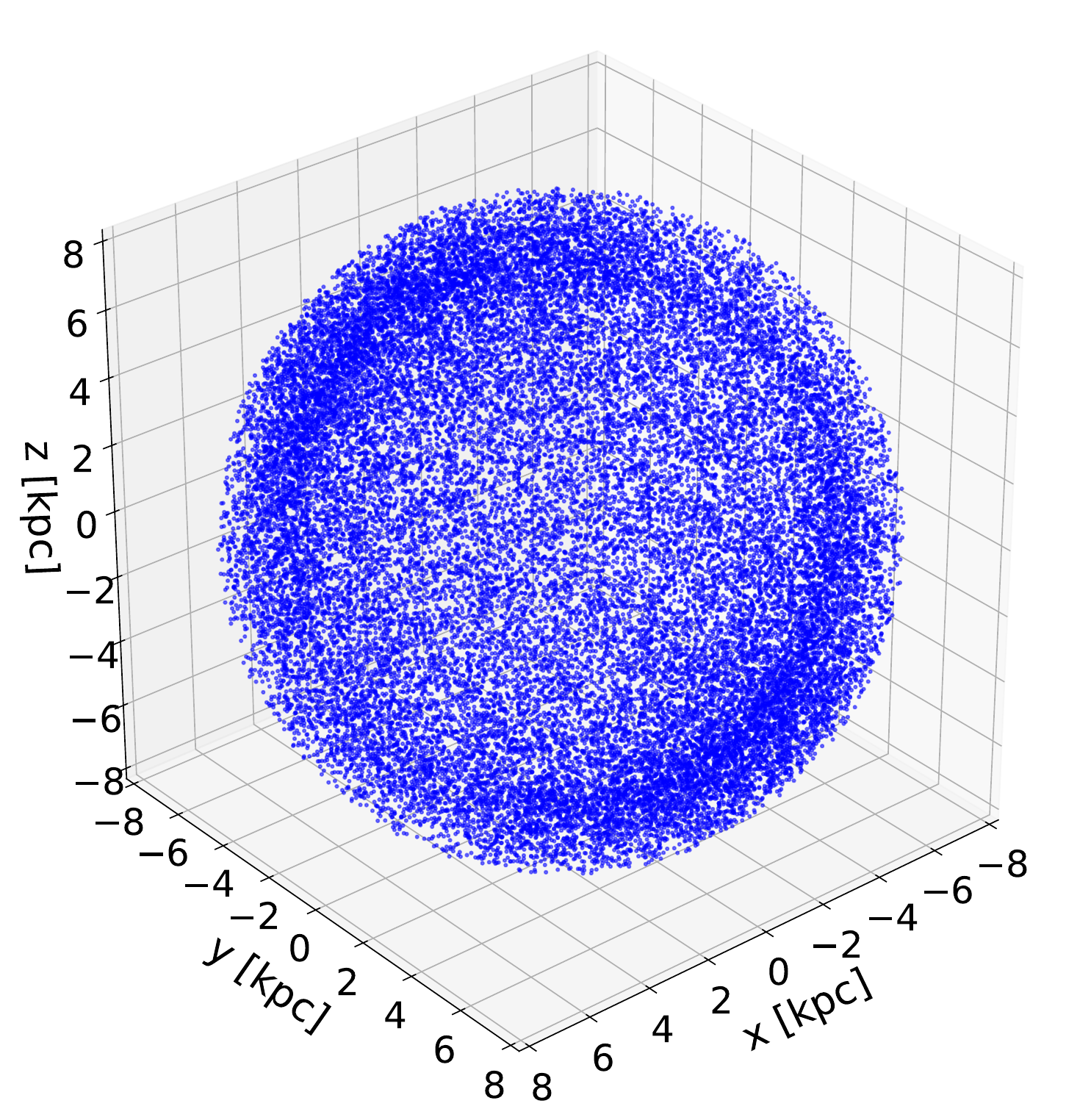}
    \caption{Method C}
    \label{fig:method_compare_3d_C}
\end{subfigure}

\caption{Three-dimensional distributions of the DM particles extracted by each method for a typical MW-like galaxy. In Methods A and B, which use the positions of stars and the stellar density, only a ring-like structure along the disc appears. In Method C, a spherical-shell structure is seen, indicating that it does not distinguish between DM associated with the disc and DM away from the disc plane.}
\label{fig:method_compare_3d}
\end{figure}

The three methods extract DM from different spatial regions. To illustrate this, we show the spatial distribution of the extracted DM particles for a representative MW-like galaxy in Fig.~\ref{fig:method_compare_3d}. The galaxy shown in both panels is Subhalo ID 554798 from the TNG50-1 data. Since Method C extracts DM particles at a distance of $7\text{--}9$ kpc from the galactic center, it picks up not only the disc-associated DM but also DM away from the disc plane; as shown in Fig.~\ref{fig:method_compare_3d}(\subref{fig:method_compare_3d_C}), these particles are distributed in a spherical shell. Methods A and B, by contrast, use the positions of stars rather than the DM particles: they extract the DM around stars identified as associated to the disc from the stellar density, thereby obtaining the DM in the vicinity of the disc. As shown in Fig.~\ref{fig:method_compare_3d}(\subref{fig:method_compare_3d_AB}), the halo component is largely removed for this representative galaxy, substantially reducing the contribution from regions away from the disc plane. As a result, the DM in the vicinity of the disc remains in a ring-like shape.

\section{Results: Local DM Density and Velocity Distribution}
\label{sec:results}

\subsection{Local DM Density}
\label{sec:local_dm_density}

\begin{figure}[tbp]
\centering
\begin{subfigure}{0.31\textwidth}
    \centering
    \makebox[\linewidth][c]{\includegraphics[height=0.17\textheight,keepaspectratio]{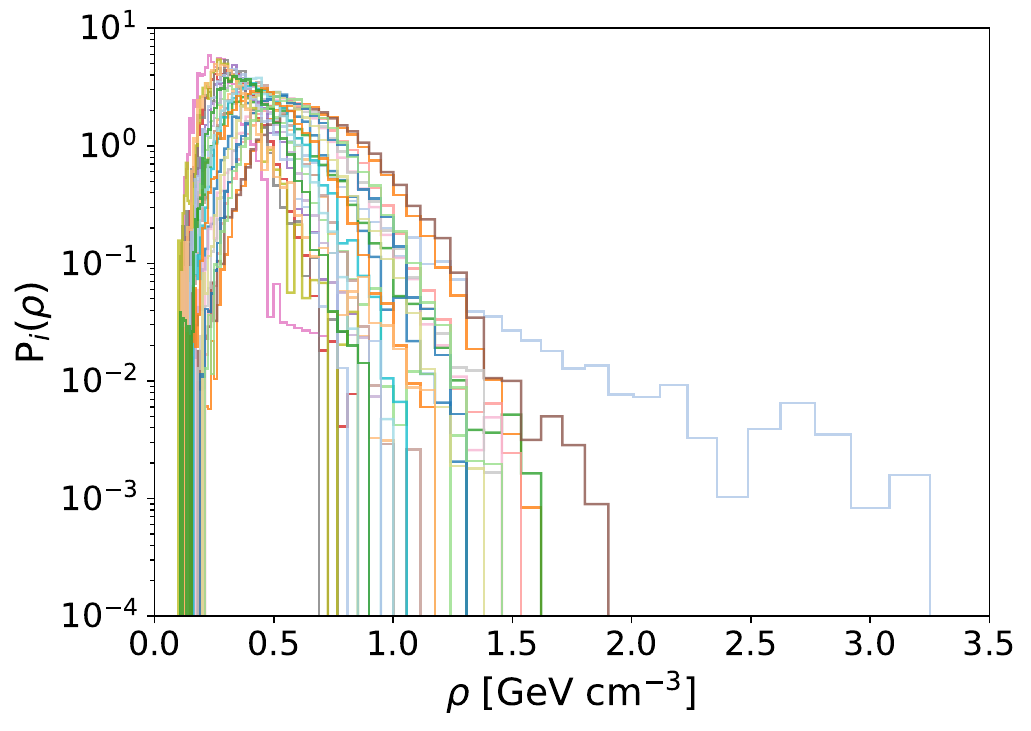}}
    \caption{Method A}
    \label{fig:rho_A}
\end{subfigure}
\hfill
\begin{subfigure}{0.31\textwidth}
    \centering
    \makebox[\linewidth][c]{\includegraphics[height=0.17\textheight,keepaspectratio]{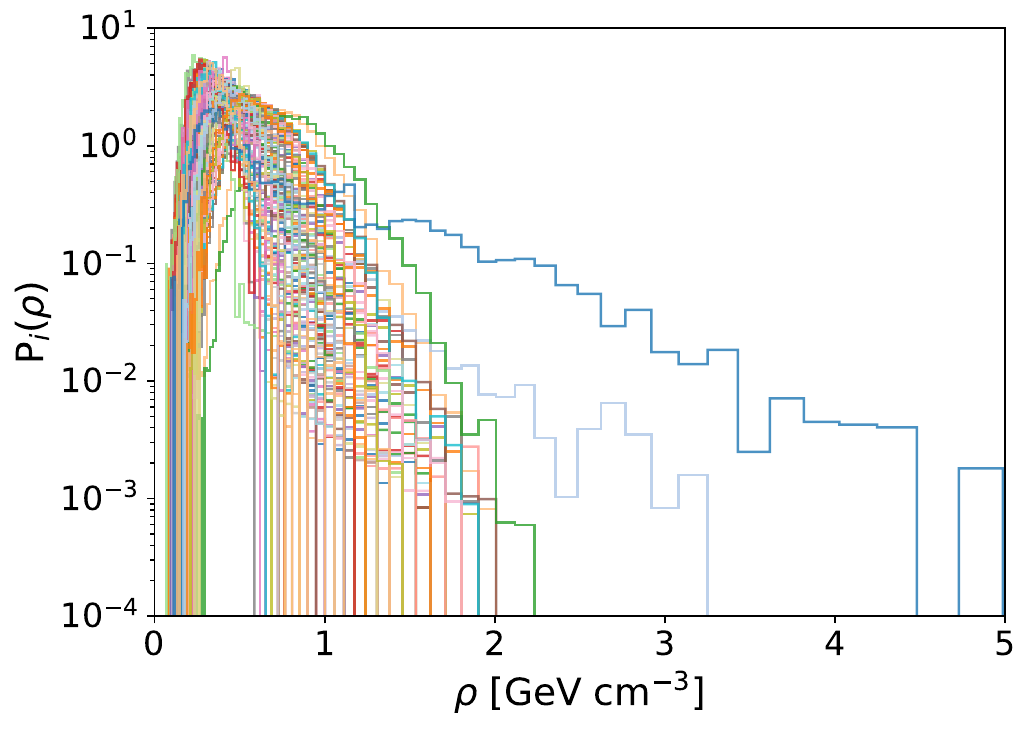}}
    \caption{Method B}
    \label{fig:rho_B}
\end{subfigure}
\hfill
\begin{subfigure}{0.31\textwidth}
    \centering
    \makebox[\linewidth][c]{\includegraphics[height=0.17\textheight,keepaspectratio]{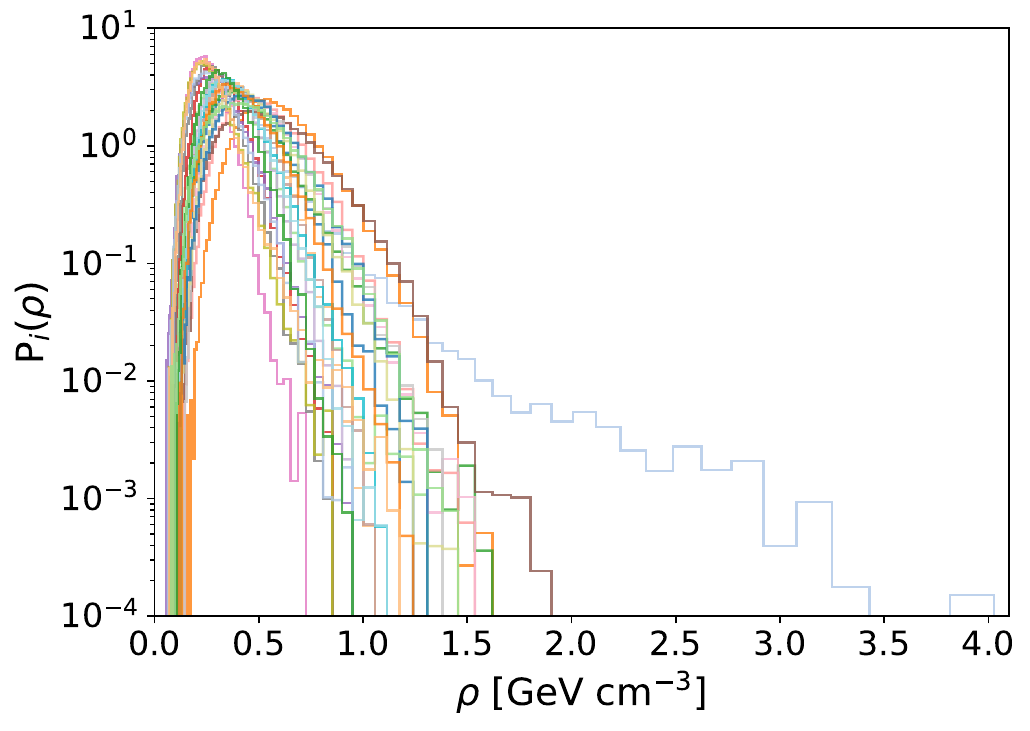}}
    \caption{Method C
    }
    \label{fig:rho_C}
\end{subfigure}
\caption{The probability distribution functions of the local DM density (DPDFs) for the MW-like galaxies extracted by each method, shown as a function of the local DM density $\rho$. Here, $P_i(\rho)$ denotes the DPDF for the $i$-th galaxy.}
\label{fig:rho}
\end{figure}

\begin{figure}[tbp]
\centering

\begin{subfigure}{0.3\textwidth}
    \centering
    \makebox[\linewidth][c]{\includegraphics[height=0.17\textheight,keepaspectratio]{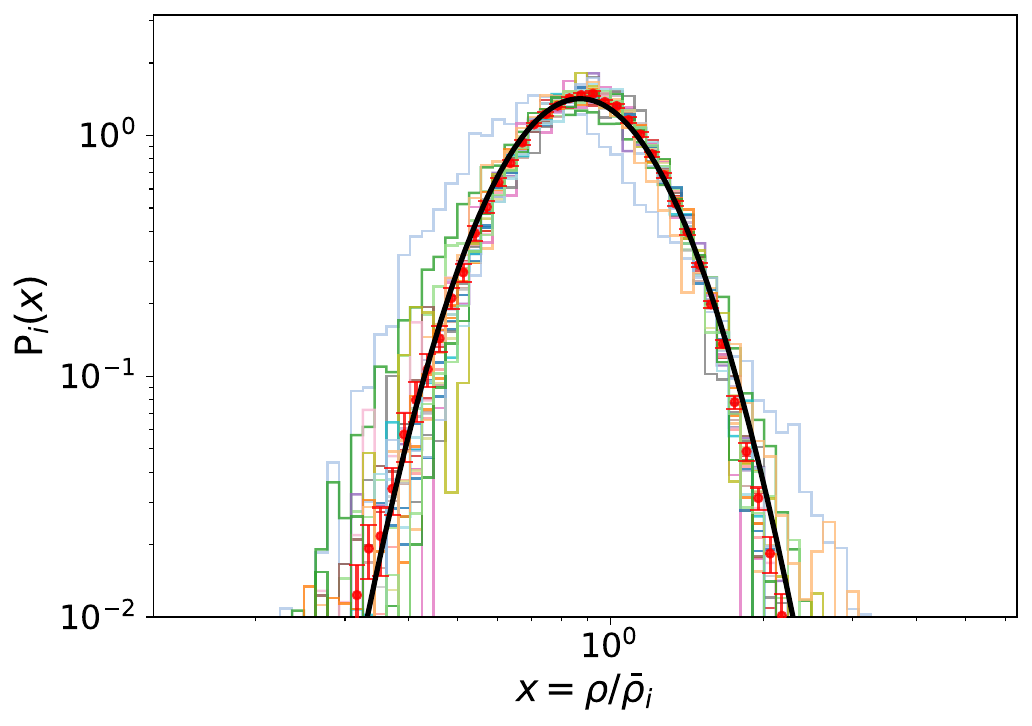}}
    \caption{Method A}
    \label{fig:dpdf_A}
\end{subfigure}
\hfill
\begin{subfigure}{0.3\textwidth}
    \centering
    \makebox[\linewidth][c]{\includegraphics[height=0.17\textheight,keepaspectratio]{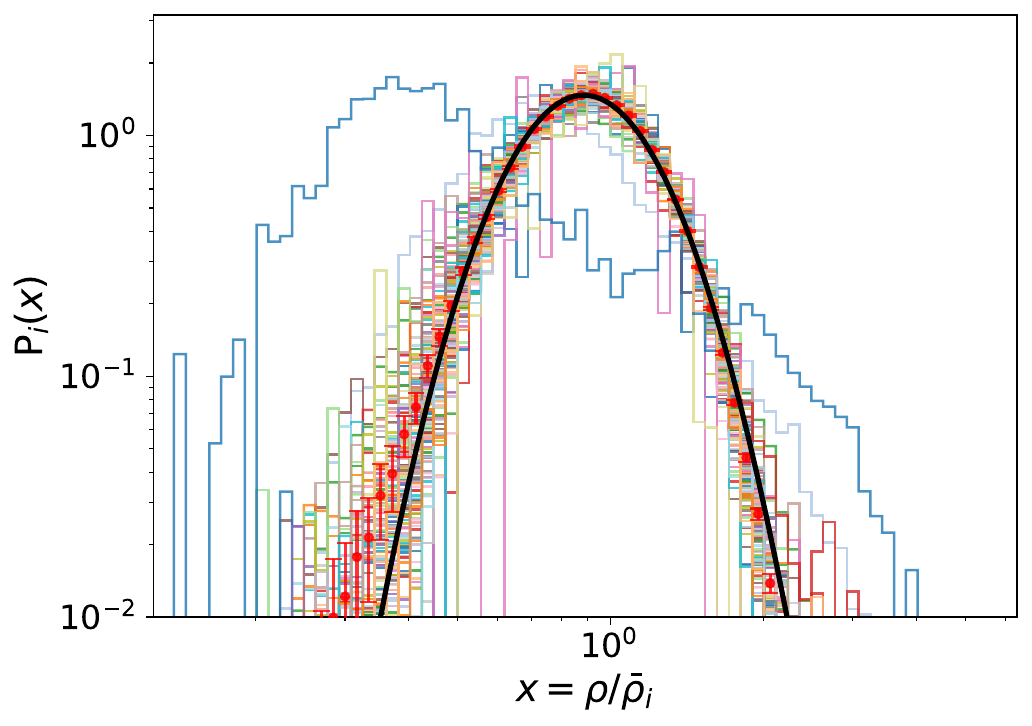}}
    \caption{Method B}
    \label{fig:dpdf_B}
\end{subfigure}
\hfill
\begin{subfigure}{0.3\textwidth}
    \centering
    \makebox[\linewidth][c]{\includegraphics[height=0.17\textheight,keepaspectratio]{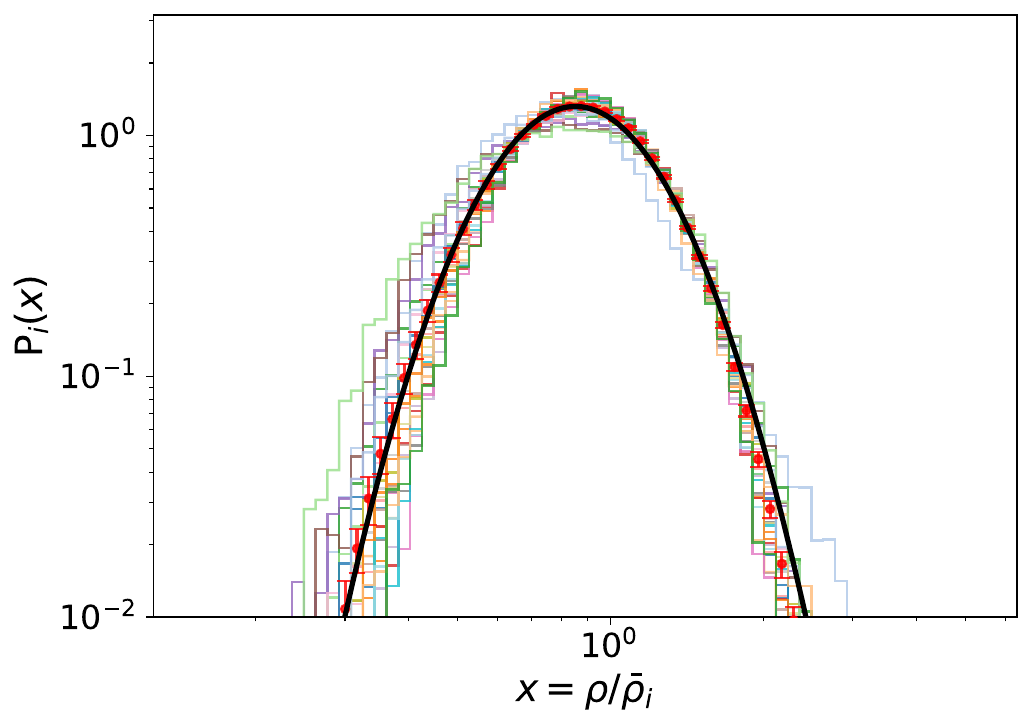}}
    \caption{Method C}
    \label{fig:dpdf_C}
\end{subfigure}
\caption{
Same as Fig.~\ref{fig:rho}, but as a function of the local DM density normalized to the mean density in each galaxy, $x = \rho/\bar\rho_i$, where $\bar{\rho}_i$ is the mean local DM density of the $i$-th galaxy. Here, $P_i(x)$ denotes the normalized DPDF of the $i$-th galaxy. The red dots show the galaxy-averaged DPDF, $\langle P(x)\rangle = \sum_i P_i(x)/N_\mathrm{gal}$, and the error bars represent the standard errors estimated from the galaxy-to-galaxy scatter. The black solid lines represent $P^\mathrm{fit}(x)$, the log-normal distribution fitted to $\langle P(x)\rangle$.}
\label{fig:dpdf}
\end{figure}

We evaluate the mass density distribution of the DM particles in the solar-neighborhood-like region extracted by the three methods discussed in Sec.~\ref{sec:method}. The probability density functions (PDFs) of the local DM density, hereafter referred to as density PDFs (DPDFs), are shown in Figs.~\ref{fig:rho} and \ref{fig:dpdf}. We estimate each DPDF as the normalized histogram of the extracted DM particles, and use it as an estimator of the underlying DPDF. In Fig.~\ref{fig:rho}, the DPDFs are shown as a function of the mass density $\rho$. For all three methods, the peak of each distribution, corresponding to the most probable local DM density, differs from galaxy to galaxy, indicating that the local DM density varies among MW-like galaxies. For a quantitative comparison, we first compute the mean local density $\bar\rho_i$ for the $i$-th galaxy by averaging over the extracted DM particles. We then average $\bar\rho_i$ over the selected galaxies to obtain $\langle \bar\rho\rangle=\sum_i \bar\rho_i/N_{\rm gal}$. Table~\ref{tab:method_comparison} summarizes the averages of the mean densities for the three methods, together with the standard deviation $\sigma_{\bar\rho}$. The errors on $\langle \bar\rho\rangle$ are the standard errors computed as $\sigma_{\bar\rho}/\sqrt{N_{\rm gal}}$. The averaged mean densities $\langle \bar\rho\rangle$ obtained by Methods A--C are all around $0.4\text{--}0.5~{\rm GeV\,cm^{-3}}$, and consistent with the observational estimates of the local DM density~\cite{Read:2014qva, deSalas:2020hbh}. Although the difference between the mean values is smaller than the galaxy-to-galaxy scatter, Method A, which selects DM in the vicinity of disc stars identified using the stellar-density criterion, gives a mean density approximately $18\%$ higher than that obtained with Method C, which selects DM solely by galactocentric distance within a spherical shell. This is consistent with the tendency reported in Refs.~\cite{Bozorgnia:2016ogo, Pato:2010yq} that the local DM density near the disc plane is higher than the spherical-shell average.

\begin{table}[tbp]
  \centering
  \caption{Average of the mean local DM densities $\langle \bar\rho\rangle$ with the standard errors, standard deviation $\sigma_{\bar\rho}$, and average peak density $\langle \rho_{\rm peak}\rangle$ with the standard errors in ${\rm GeV}\,{\rm cm}^{-3}$ for each method. The standard errors are estimated from the corresponding galaxy-to-galaxy scatter divided by $\sqrt{N_{\rm gal}}$.}

  \begin{tabular}{|c|c|c|c|}
    \hline
     & $\langle\bar\rho\rangle$ & $\sigma_{\bar\rho}$ & $\langle \rho_{\rm peak}\rangle$ \\
    \hline
    Method A & $0.47\pm0.02$  & $0.11$ & $0.41\pm0.02$ \\
    \hline
    Method B & $0.48\pm0.01$  & $0.11$ & $0.42\pm0.01$ \\
    \hline
    Method C & $0.40\pm0.02$  & $0.09$ & $0.34\pm0.02$ \\
    \hline
  \end{tabular}
  \label{tab:method_comparison}
\end{table}

Figure~\ref{fig:dpdf} shows the DPDFs as a function of the density normalized to the mean density of each galaxy $\bar\rho_i$. 
In Method A, the normalized DPDFs are similar among the selected MW-like galaxies, suggesting that, once the mean local DM density of a galaxy is specified, the probability distribution of its local density can be inferred from the common normalized DPDF. Since dynamical estimates based on stellar motions effectively probe the DM density averaged over a finite region around the Sun \cite{Read:2014qva}, the observationally inferred local DM density is expected to be close to the mean density $\bar{\rho}_i$ used here.
The DPDFs in Method B are considerably more scattered than in Method A.
This reflects the more diverse set of galaxies selected by relaxing the criteria for MW-like galaxies in Method B. For example, the galaxy presented by the blue line in Fig.~\ref{fig:dpdf}(\subref{fig:dpdf_B}) departs from the bulk of the distributions. We found that the galaxy has a nearby companion, suggesting that the local DM density distribution is affected by the neighboring galaxy. Method C, which uses the same galaxy sample as Method A but adopts a distance-based DM extraction, also gives normalized DPDFs that are similar among the galaxies. 

To quantify the differences among the three methods, we fit a log-normal distribution to the galaxy-averaged normalized DPDF for each method. The log-normal distribution used here is given by
\begin{equation}
  P^{\mathrm{fit}}(x) = \frac{\mathcal{A}} {x\sigma\sqrt{2\pi}} \exp\left[ -\frac{(\ln x-\mu)^2}{2\sigma^2} \right],
\label{eq:lognormal}
\end{equation}
where $\mathcal{A}$ is the normalization factor, $x=\rho/\bar{\rho}_i$ is the dimensionless local DM density normalized by the mean density of the $i$-th galaxy, and $\mu$ and $\sigma$ are the location and width parameters in $\ln x$, respectively. For each method, we fit Eq.~\eqref{eq:lognormal} to the galaxy-averaged normalized DPDF in logarithmic space. The standard errors estimated from the galaxy-to-galaxy scatter are converted to the corresponding uncertainties in logarithmic space and used as weights in the fit.
The widths obtained from the fit to the DPDFs of $\rho/\bar\rho_i$ are $\sigma =0.31\pm 0.02$, $0.29\pm 0.01$, and $0.33\pm 0.01$ for Methods A, B and C, respectively, where the uncertainties are estimated using galaxy-level bootstrap resampling. The central value of the width is larger in Method C than in Method A, although the difference does not exceed the errors. The averaged mean density is $0.47\pm 0.02~{\rm GeV\,cm^{-3}}$ in Method A and $0.40\pm 0.02~{\rm GeV\,cm^{-3}}$ in Method C, a difference that exceeds the errors. The lower averaged mean density in Method C is likely because Method C includes DM particles in lower-density environments away from the disc plane. We also fit the normalized DPDF $P_i(x)$ of each galaxy separately with the log-normal distribution in Eq.~\eqref{eq:lognormal}. For the $i$-th galaxy, the peak of the fitted distribution $P_i^{\mathrm{fit}}(x)$ is given by $x_{\mathrm{peak},i}=\exp(\mu_i-\sigma_i^2)$, where $\mu_i$ and $\sigma_i$ are the best-fit parameters for that galaxy. The corresponding peak density is then obtained as $\rho_{\mathrm{peak},i}=\bar{\rho}_i x_{\mathrm{peak},i}$. We define the average peak density as $\langle \rho_{\mathrm{peak}}\rangle=\sum_i \rho_{\mathrm{peak},i}/N_{\mathrm{gal}}$. The resulting values are $0.41\pm 0.02$, $0.42\pm 0.01$, and $0.34\pm 0.02\,{\rm GeV\,cm^{-3}}$ for Methods A, B, and C, respectively. The uncertainties are standard errors estimated from the galaxy-to-galaxy scatter. The difference in the peak density between Methods A and C is about $20\%$, which is smaller than the overall uncertainty in the observationally inferred local DM density \cite{Read:2014qva, deSalas:2020hbh}. Its impact on the interpretation of direct detection experiments is therefore limited.

\subsection{Velocity Distribution}
\label{sec:velocity_distribution}

\begin{figure}[tbp]
\centering
\begin{subfigure}{0.32\textwidth}
    \centering
    \includegraphics[width=\linewidth]{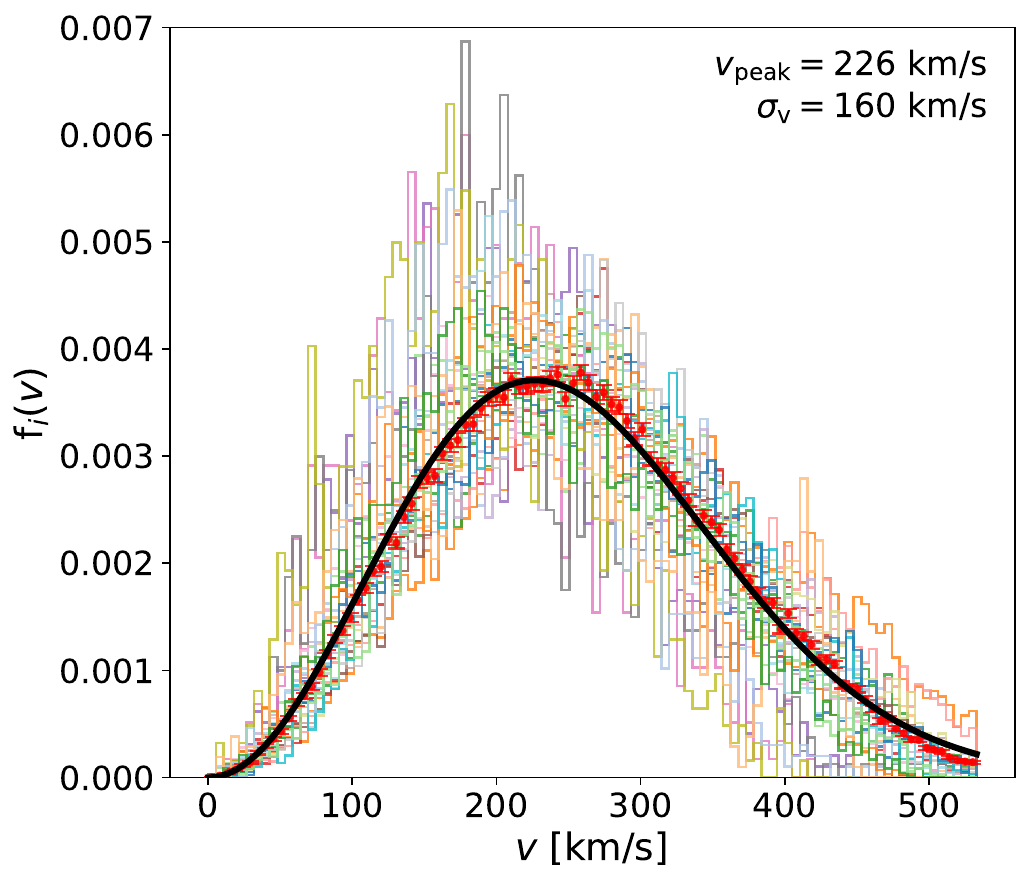}
    \caption{Method A}
\end{subfigure}
\hfill
\begin{subfigure}{0.32\textwidth}
    \centering
    \includegraphics[width=\linewidth]{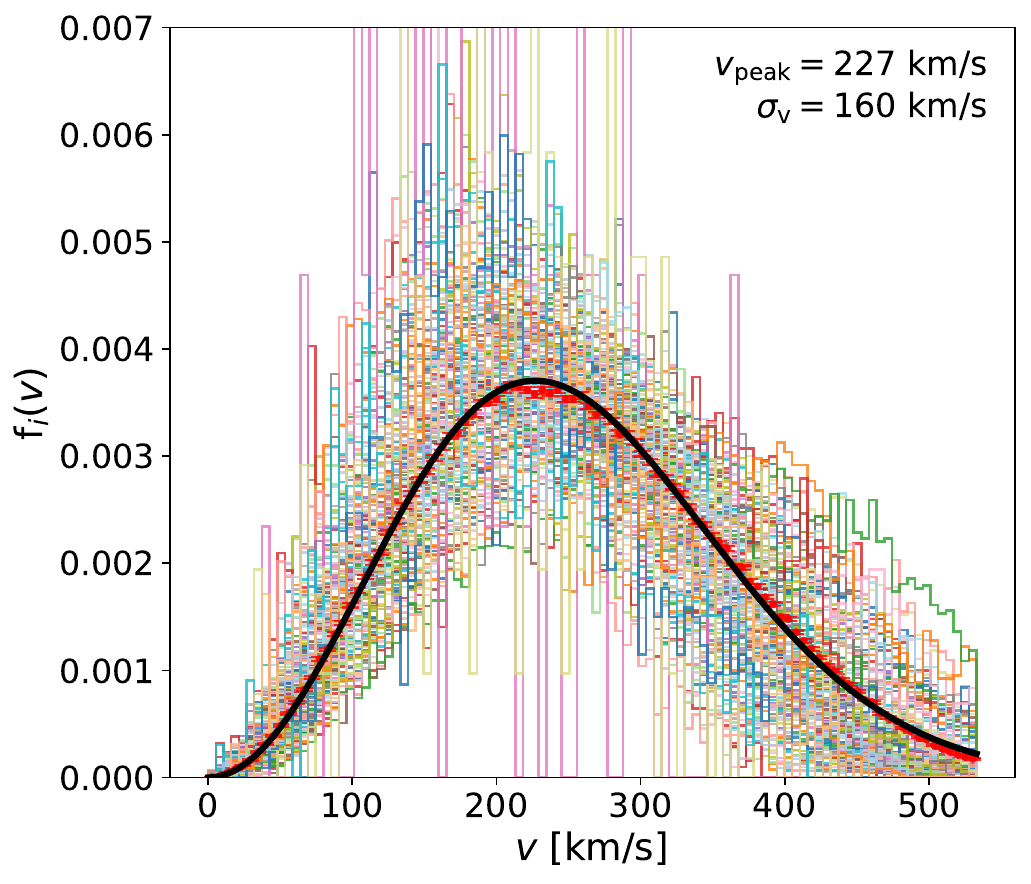}
    \caption{Method B}
\end{subfigure}
\hfill
\begin{subfigure}{0.32\textwidth}
    \centering
    \includegraphics[width=\linewidth]{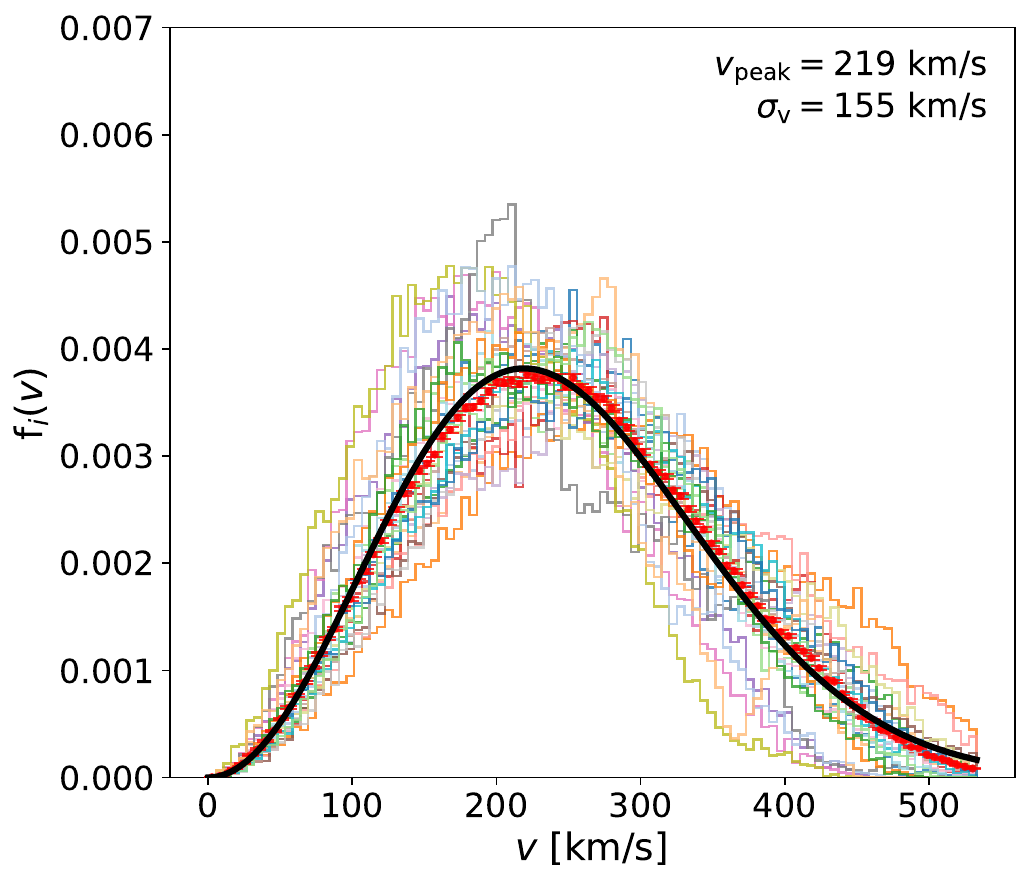}
    \caption{Method C}
\end{subfigure}

\caption{DM velocity distributions in the solar-neighborhood-like regions of MW-like galaxies extracted by each method. The thin colored lines show $f_i(v)$, the normalized histogram of the velocity for the $i$-th galaxy. The red dots show $f_{\mathrm{all}}(v)$, obtained by combining the extracted DM particles from all the galaxies, and the error bars indicate the corresponding Poisson errors. The black solid lines show $f^{\mathrm{fit}}(v)$, the Maxwellian distributions fitted to $f_{\mathrm{all}}(v)$.}
\label{fig:vel_all}
\end{figure}

\begin{figure}[tbp]
\centering

\begin{subfigure}{0.32\textwidth}
    \centering
    \includegraphics[width=\linewidth]{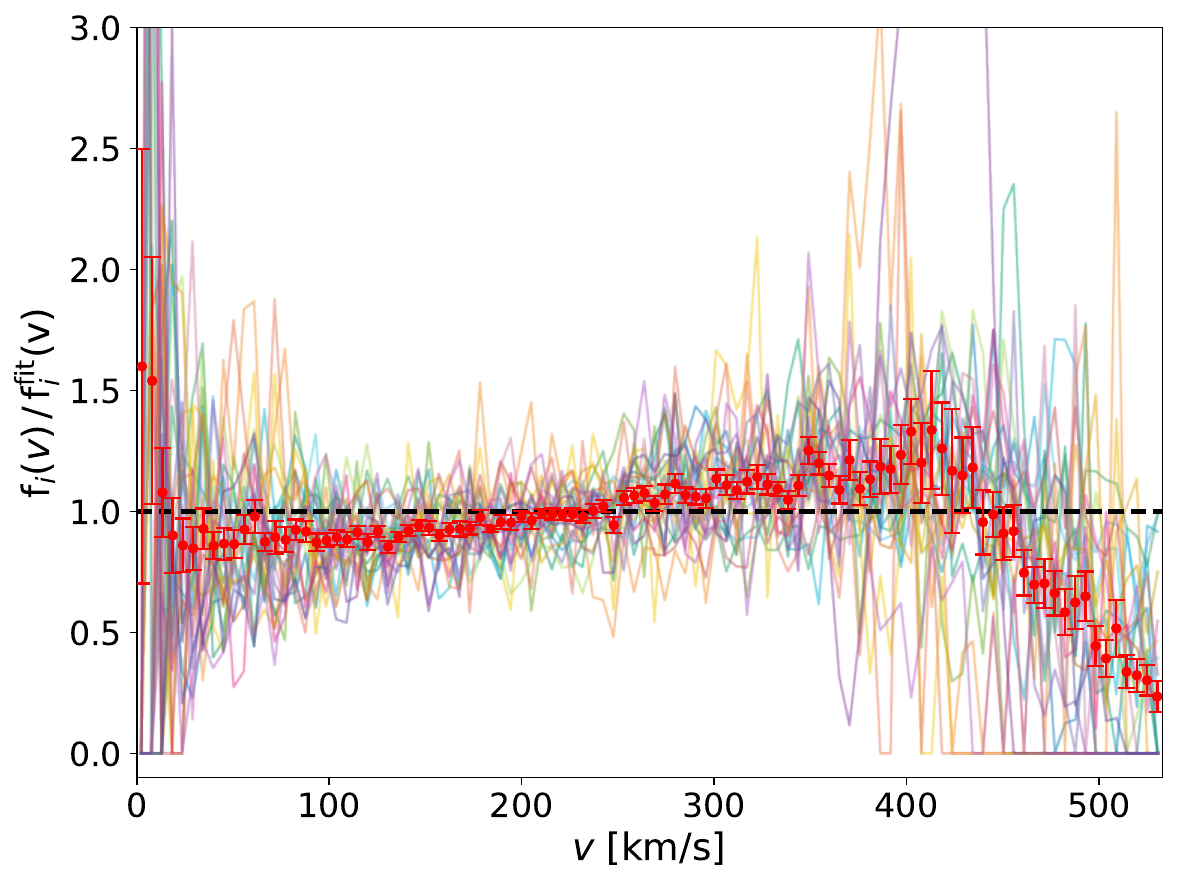}
    \caption{Method A}
\end{subfigure}
\hfill
\begin{subfigure}{0.32\textwidth}
    \centering
    \includegraphics[width=\linewidth]{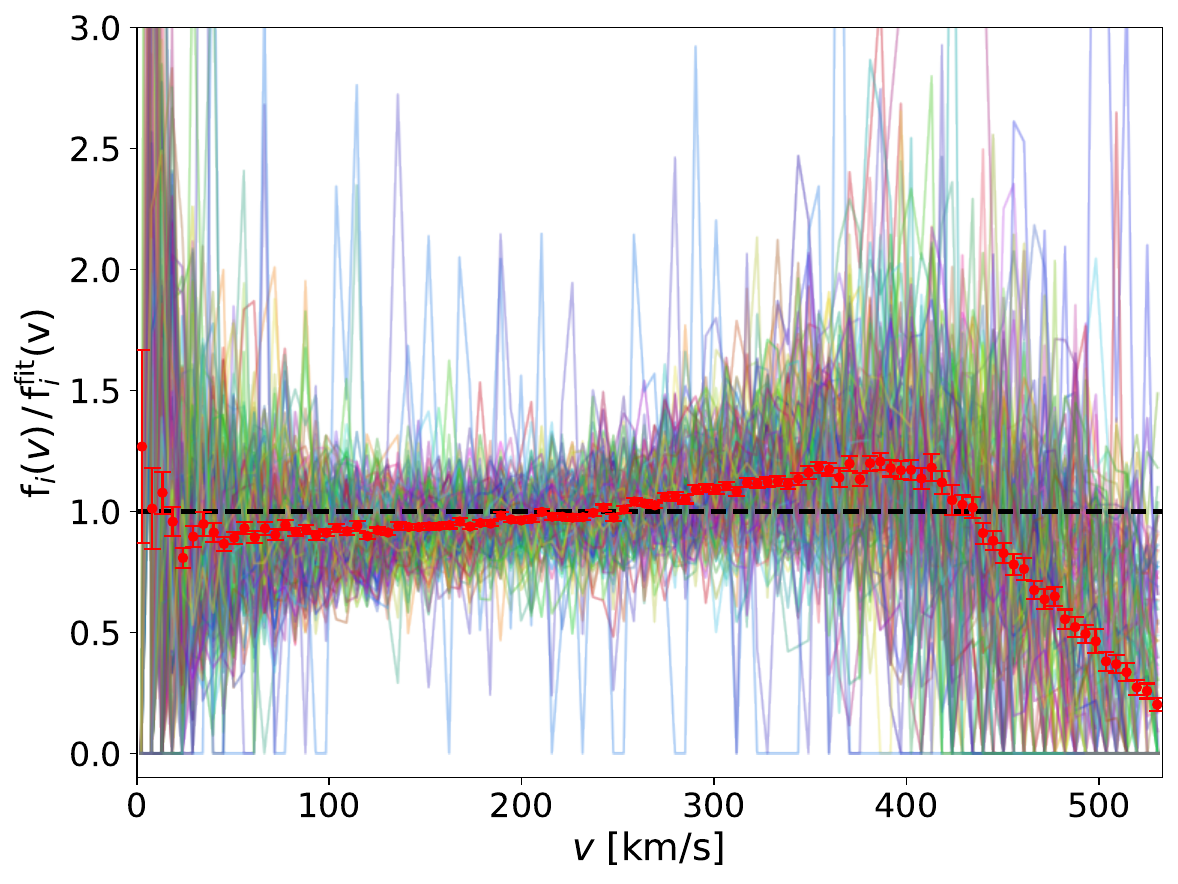}
    \caption{Method B}
\end{subfigure}
\hfill
\begin{subfigure}{0.32\textwidth}
    \centering
    \includegraphics[width=\linewidth]{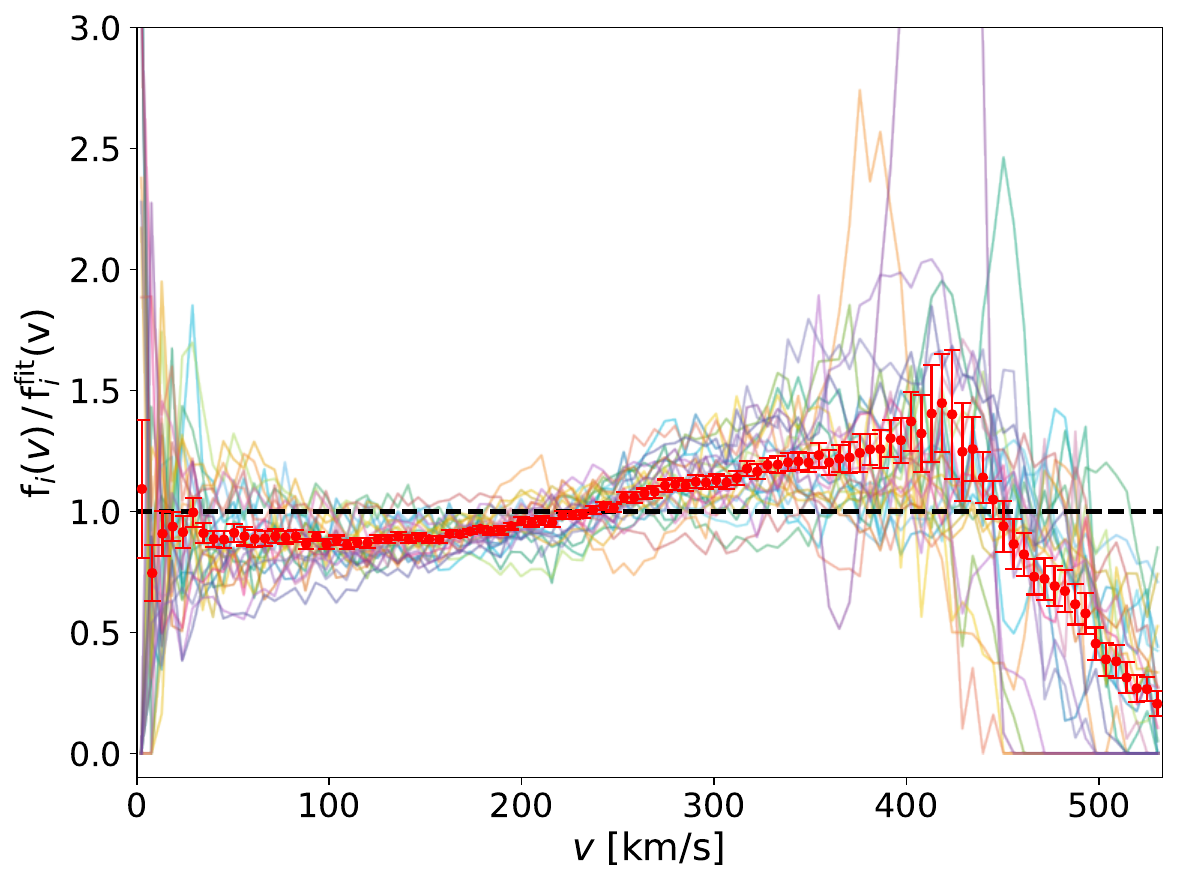}
    \caption{Method C}
\end{subfigure}

\caption{Ratios $f_i(v)/f_i^{\mathrm{fit}}(v)$ of the DM velocity distributions to their individually fitted Maxwellian distributions in the solar-neighborhood-like regions. The colored lines show the ratios for individual galaxies, where $f_i^{\mathrm{fit}}(v)$ is obtained by fitting the Maxwellian distribution separately to $f_i(v)$. The red dots show the galaxy-averaged ratio, where each galaxy is given equal weight, and the error bars represent the standard errors estimated from the galaxy-to-galaxy scatter. The black dashed line indicates a ratio of unity.}
\label{fig:vel_ratio}
\end{figure}
Figure~\ref{fig:vel_all} shows the DM velocity distributions in the solar-neighborhood-like regions of the selected galaxies, together with their combined distribution. The thin colored lines show $f_i(v)$, the normalized histogram of the velocity for the $i$-th galaxy. The red dots show $f_{\rm all}(v)$, the normalized histogram obtained by combining the extracted DM particles from all the selected galaxies, where the count in each velocity bin is normalized by the total number of particles and the bin width. The black solid lines show $f^{\rm fit}(v)$, the Maxwellian distributions fitted to $f_{\rm all}(v)$. In what follows, we treat these normalized histograms as estimators of the underlying velocity distribution functions. The error bars are estimated from the Poisson uncertainty $\sqrt{N_{j}}$ of the number of DM particles $N_{j}$ in the $j$-th velocity bin, normalized by the bin width $\Delta v$ and the total number of particles $N_{\rm tot}$ within $0\leq v\leq v_{\rm esc}$ summed over all the selected galaxies, as $\sigma_{{\rm PDF}, j} = \sqrt{N_{j}}/(N_{\rm tot}\,\Delta v)$. We fit $f_{\mathrm{all}}(v)$ with the isotropic Maxwellian distribution
\begin{equation}
f^\mathrm{fit}(v) = \frac{4\pi v^2}{N} \left(\frac{1}{2\pi\sigma_v^2}\right) ^ {3/2} \exp \left(-\frac{v^2}{2\sigma_v^2}\right) \qquad
(0\leq v\leq v_{\rm esc}),  
\label{eq:maxwell}
\end{equation}
where $v$ is the magnitude of the DM particle velocity relative to the bulk velocity of the host subhalo, and $\sigma_v$ is the scale parameter of the Maxwellian distribution.
The normalization constant $N$ is given by
\begin{equation}
N = \int_0^{v_{\rm esc}} 4\pi v^2 \left(\frac{1}{2\pi\sigma_v^2}\right) ^ {3/2} \exp \left(-\frac{v^2}{2\sigma_v^2}\right) dv  
\label{eq:normalization_constant}
\end{equation}
We use only particles with $v\leq v_{\rm esc}$, adopting $v_{\rm esc}=533~\mathrm{km\,s^{-1}}$ \cite{Piffl:2013mla}. The parameter $\sigma_v$ is determined by fitting the Maxwellian distribution to the binned PDF, using the Poisson errors as weights. For each bin, the model prediction is calculated by integrating the Maxwellian distribution over the bin range and dividing it by the bin width. The peak velocities obtained from the Maxwellian fits are $v_{\mathrm{peak}}=\sqrt{2}\sigma_v = $ $226.4\pm 0.3$, $226.7\pm 0.1$, and $219.1\pm 0.1~\mathrm{km\,s^{-1}}$ for Methods A, B, and C, respectively. The uncertainties in $v_{\mathrm{peak}}$ are obtained by propagating the uncertainties in $\sigma_v$ derived from the covariance matrix of the fit. In all cases, these values are close to the standard halo model (SHM) value of $220~\mathrm{km\,s^{-1}}$~\cite{Evans:2018bqy}. The shape of the velocity distribution shows no major dependence on the method and is well approximated by a Maxwellian in every case. Moreover, although Methods A and C differ in the extraction of the solar-neighborhood-like region, with Method C additionally including DM away from the disc plane, their velocity distributions show no substantial difference. This suggests that, at least in the present analysis, restricting the extraction to DM associated with the disc has little effect on the velocity distribution. 

Figure~\ref{fig:vel_ratio} shows the ratio of each galaxy's velocity distribution to its best-fit Maxwellian distribution, $f_i(v)/f^{\rm fit}_i(v)$, where the Maxwellian distribution is fitted separately to each galaxy. 
The galaxy-averaged ratio remains close to unity around the peak of the velocity distribution, where most of the DM particles lie, while it tends to fall below unity at the high- and low-velocity ends, except for the lowest bins where the errors are large.
For traditional WIMPs, the event rate is dominated by DM particles with velocities near the peak, where the distribution is well described by a Maxwellian, so the deviations at the low- and high-velocity ends are expected to have only a limited impact on the event rate in direct detection experiments. However, for light DM or for experiments with a high recoil-energy threshold, the event rate is mainly determined by the high-velocity tail. In this case, the deviations from a Maxwellian may have a larger impact on the event rate.

\begin{figure}[tbp]
  \centering
  \includegraphics[width=0.4\linewidth]{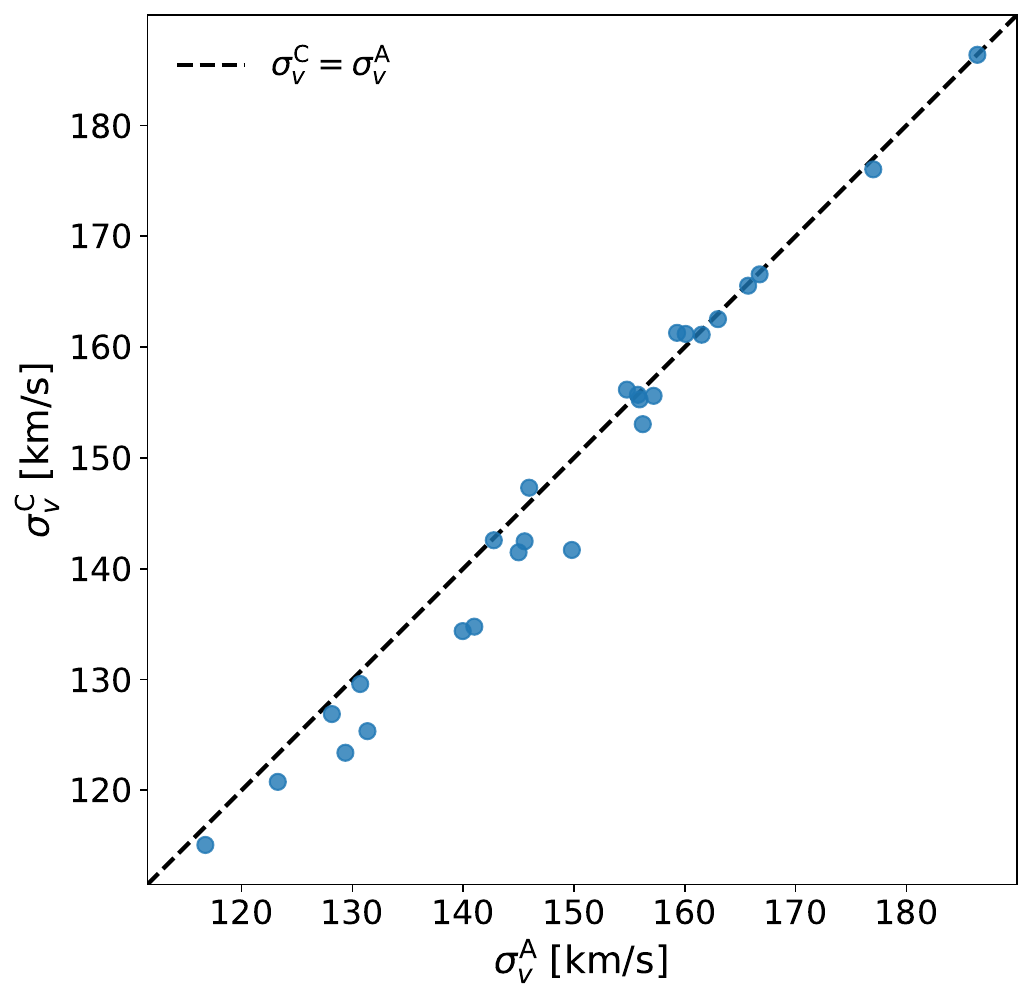}

  \caption{Comparison of the Maxwellian parameters $\sigma_v^{\mathrm{A}}$ and $\sigma_v^{\mathrm{C}}$ obtained for individual galaxies in Methods A and C, respectively. Each blue point represents the results obtained for the same galaxy in the two methods. The black dashed line indicates $\sigma_v^{\mathrm{A}}=\sigma_v^{\mathrm{C}}$.}

  \label{fig:sigmav_AC}
\end{figure}

Going beyond the averages discussed so far, we now examine Methods A and C galaxy by galaxy. Fig.~\ref{fig:sigmav_AC} compares the Maxwellian parameters $\sigma_v$ obtained for the 26 galaxies common to Methods A and C. The values obtained with the two methods are strongly correlated and lie close to the line $\sigma_v^{\mathrm{A}}=\sigma_v^{\mathrm{C}}$. However, $\sigma_v^{\mathrm{A}}$ is larger than $\sigma_v^{\mathrm{C}}$ for 22 of the 26 galaxies. The mean values of $\sigma_v$ are $149.58\pm 3.28\ {\rm km\,s^{-1}}$ and $147.77\pm 3.54\ {\rm km\,s^{-1}}$ for Methods A and C, respectively. The mean difference is $\langle \sigma_v^{\mathrm{A}} - \sigma_v^{\mathrm{C}}\rangle = 1.81\pm 0.53\ {\rm km\,s^{-1}}$. The uncertainties are standard errors estimated from the galaxy-to-galaxy scatter. Thus, although the difference is small, Method A tends to yield slightly larger Maxwellian parameters than Method C.

\subsection{Comparison with Previous Studies}
\label{sec:comparison_previous_studies}

In previous simulation studies, the local DM density on the galactic disc tends to be higher than the average density in a spherical shell that includes the halo \cite{Pato:2010yq, Bozorgnia:2016ogo}. Consistent with previous simulation studies, we find a higher local DM density in Method A than in Method C, although the difference is small compared with the range of values reported in previous studies. The local DM density reported in previous studies is $0.3\text{--}0.7~{\rm GeV\,cm^{-3}}$ \cite{Ling:2009eh, Bozorgnia:2016ogo, Staudt:2024tdq, Folsom:2025lly}, lying in the upper part of the range suggested by observations \cite{deSalas:2020hbh}. The local DM density obtained in this study is also slightly high, similar to these studies.

The velocity distribution of the local DM has also been estimated from cosmological simulations. In simulations including baryons, the velocity distribution is reported to be Maxwellian, and it is pointed out that the gravitational potential of the baryonic disc increases the amount of high-velocity DM and brings the velocity distribution closer to a Maxwellian than in DM-only simulations \cite{Kelso:2016qqj, Sloane:2016kyi, Bozorgnia:2017brl, Santos-Santos:2023ubx, Shpigel:2025ulk, Zhang:2026qnl}. Deviations from a Maxwellian are, however, also reported even in a simulation including baryons, where the high-velocity tail drops more sharply \cite{Ling:2009eh}. The velocity distribution obtained in this study is well described by a Maxwellian, consistent with the effect of baryons discussed above. The peak velocities are also comparable to those in previous studies \cite{Folsom:2025lly}.

\section{Conclusion}
\label{sec:conclusion}

In this study, using the TNG50-1 data of IllustrisTNG, we select MW-like galaxies and the solar-neighborhood-like regions within them realistically, and evaluate the mass density and velocity distribution of DM.
For the selection of MW-like galaxies, the selection criteria are that the halo mass of the galaxy and the bulge-to-total stellar mass ratio $B/T$ are close to those of the MW. For the extraction of the solar-neighborhood-like regions, we obtain the DM particles associated with the disc by using those in the vicinity of stars located at distances of $7\text{--}9$ kpc from the galactic center. We further adopt a prescription that uses the stellar density distribution to select the disc stars, which suppresses contamination by halo DM. As a result, compared to the method that treats the DM at $7\text{--}9$ kpc from the galactic center as DM in the solar neighborhood, we develop an analysis method that extracts the DM associated with the galactic disc more realistically.

We find that the DPDFs as a function of the normalized DM density in the solar-neighborhood-like regions in the 26 galaxies obtained under the stringent selection criteria show very little variation among the galaxies (Fig.~\ref{fig:dpdf}). In contrast, relaxing the selection criteria for MW-like galaxies yields large variations in the shape of the distribution and a more diverse set of galaxies. Fitting the DPDFs with log-normal distributions, we obtain a width of $0.31\pm 0.02$ for the former and $0.33\pm 0.01$ for the latter. The peak density, on the other hand, is higher for the disc-based extraction, $0.41\pm0.02\,{\rm GeV\,cm^{-3}}$, than for the distance-based one, $0.34\pm0.02\,{\rm GeV\,cm^{-3}}$. As expected, the disc-based extraction yields differences in the estimated parameters, both in the local DM density and in the velocity distribution. These differences, however, are at most about 10--20\% across the methods examined here. The local DM density of $0.47\pm 0.02~{\rm GeV\,cm^{-3}}$ obtained under the stringent selection criteria is consistent with the range estimated from observations, lying in the upper part of that range. 
The local DM density obtained under the less stringent selection criteria is also within $0.4\text{--}0.5~{\rm GeV\,cm^{-3}}$, showing no substantial difference between the selection methods for the MW-like galaxies and the solar-neighborhood-like regions. The DM velocity distribution depends little on the selection. The peak velocities obtained with the three methods are close to the characteristic value adopted in the SHM. Thus, while the local DM density depends slightly on the extraction method, the shape of the DPDF and the velocity distribution are largely insensitive to it.

Evaluating the impact of these results on direct detection experiments, the local DM density obtained by the method of this study is about $0.47~{\rm GeV\,cm^{-3}}$, which is slightly higher than the $0.3~{\rm GeV\,cm^{-3}}$ assumed in direct detection experiments, but it is within the range of uncertainty evaluated in previous studies. Also, the velocity distribution can be well approximated by an isotropic Maxwellian distribution, with a peak velocity close to the characteristic value adopted in the SHM. 
Therefore, the exclusion curves of direct detection experiments based on the density and velocity distribution obtained in this study would differ little from the limits reported by each experiment. We fitted the velocity distribution with an isotropic Maxwellian in this study. However, in realistic galactic discs, the velocity dispersions may differ among the three spatial components owing to the structure and dynamical evolution of the galaxy \cite{Read:2014qva,deSalas:2020hbh, Bozorgnia:2016ogo}. A component-wise analysis of such an anisotropic velocity distribution could change the impact on direct detection experiments. In that case, the disc-extraction method proposed in this study, which suppresses the halo contribution and preferentially selects DM associated with the disc region, should be well suited to evaluating the disc-specific kinematics and velocity anisotropy. We leave this for future work. 

\appendix
\section{Identifying Disc Stars via Three-Gaussian Decomposition of the Stellar Density}
\label{sec:gaussian}

When extracting DM particles from a solar-neighborhood-like region, selecting particles solely by their galactocentric distance would also include DM from regions away from the disc plane. We therefore identify the solar-neighborhood-like region using stars rather than selecting DM particles directly. Specifically, we first select stars located at galactocentric distances of $7\text{--}9$ kpc and then extract DM particles in their vicinity. As shown in Fig.~\ref{fig:method_compare_3d}, this method substantially reduces the contribution from regions away from the disc plane. Even so, in some galaxies, halo stars may still be mixed in, and the halo contribution cannot be fully removed. To handle such cases, we apply a further prescription that suppresses the halo component using the stellar density as an indicator. This Appendix describes that prescription.

\begin{figure}[tbp]
  \centering
  \includegraphics[width=0.45\linewidth]{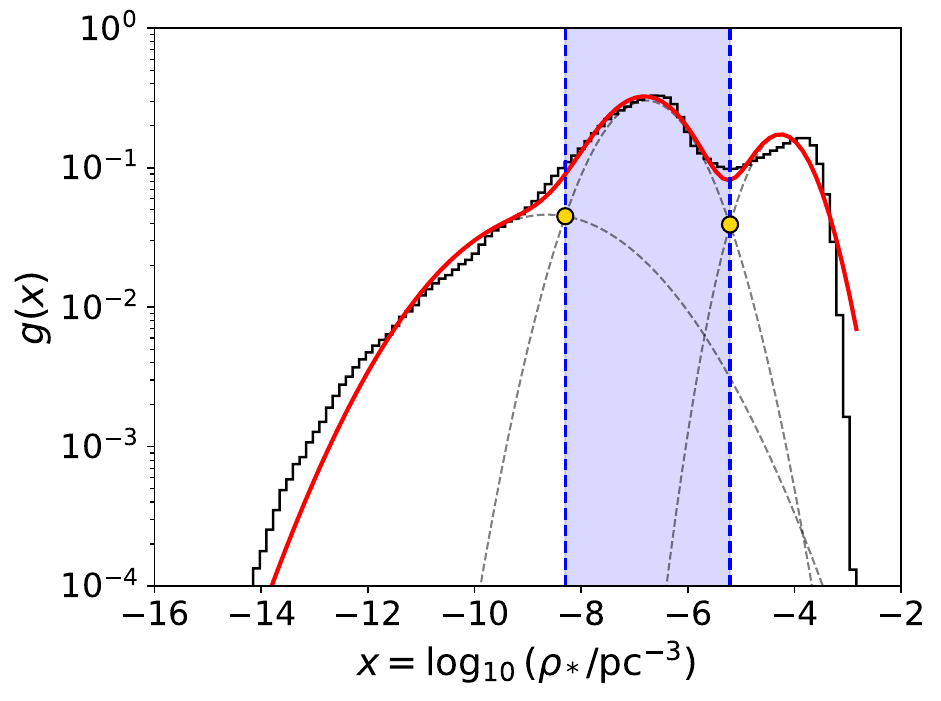}
  \caption{Histogram of the logarithmic stellar number density $x=\log_{10}(\rho_*/\mathrm{pc}^{-3})$.
  The black line shows $g(x)$, the normalized histogram of $x$, and
  the red line shows the three-Gaussian fit $g^{\mathrm{fit}}(x)$ used to classify stars into the bulge, disc, and halo components. The thin black dashed lines show the individual Gaussians. The blue-shaded region between the intersection points indicates the stars identified as the disc component.}
  \label{fig:stellar_all}
\end{figure}

For each galaxy, we decompose the logarithmic stellar-density distribution into three Gaussian components, which we associate with the halo, disc, and bulge populations. As an example, Figure~\ref{fig:stellar_all} shows the stellar-density distribution averaged over the 26 galaxies selected in Method A. The actual decomposition used for star classification is performed separately for each galaxy. For each star particle, we estimate the local stellar number density $\rho_*$ using its \texttt{StellarHsml} value as $\rho_* = 32/\left[(4\pi/3)R_{\mathrm{hsml}}^3\right]$,
where $R_{\mathrm{hsml}}$ is the physical stellar smoothing length. The probability density function $g(x)$ of the logarithmic stellar number density, $x=\log_{10}(\rho_*/\mathrm{pc}^{-3})$,
is estimated as the normalized histogram of the star particles and fitted with a superposition of three Gaussians corresponding to the halo (h), disc (d), and bulge (b) components,
\begin{equation}
g^{\mathrm{fit}}(x) = \sum_{k \in \{\mathrm{h}, \mathrm{d}, \mathrm{b}\}} \frac{w_k}{\sqrt{2\pi\sigma_k^2}} \exp \left( -\frac{(x - \mu_k)^2}{2\sigma_k^2} \right),
\label{eq:three_gaussian}
\end{equation}
where $w_k$ is the mixing coefficient of each component (with $\sum_k w_k = 1$), $\mu_k$ is the mean, and $\sigma_k^2$ is the variance. These parameters are determined by maximum-likelihood estimation using the expectation-maximization algorithm on the simulation data. The means satisfy $\mu_{\mathrm{h}} < \mu_{\mathrm{d}} < \mu_{\mathrm{b}}$, so that the lowest-density component corresponds to the halo, the intermediate one to the disc, and the highest-density one to the bulge.
The boundaries between components are defined as the intersection points at which the probability densities of two adjacent Gaussians are equal,
\begin{equation}
\frac{w_k}{\sqrt{2\pi\sigma_k^2}} \exp \left( -\frac{(x - \mu_k)^2}{2\sigma_k^2} \right)
= \frac{w_l}{\sqrt{2\pi\sigma_l^2}} \exp \left( -\frac{(x - \mu_l)^2}{2\sigma_l^2} \right).
\label{eq:intersection}
\end{equation}
We solve this for the adjacent pairs $(k,l)=(\mathrm{h},\mathrm{d})$ and $(\mathrm{d},\mathrm{b})$ to obtain the intersection points. Denoting the halo-disc intersection by $x_{\mathrm{hd}}$ and the disc-bulge intersection by $x_{\mathrm{db}}$, a star particle is identified as belonging to the disc if $x_{\mathrm{hd}}\le x\le x_{\mathrm{db}}$. In Methods A and B, these three Gaussians are partitioned at their intersection points, and the stars in the region between them (shaded in blue in Fig.~\ref{fig:stellar_all}) are identified as disc stars and used in the analysis.

\begin{figure}[tbp]
\centering
\begin{subfigure}{0.45\textwidth}
    \centering
    \includegraphics[width=\linewidth]{pos_AB_554798.pdf}
    \caption{DM particles extracted using stars at $7\text{--}9$ kpc from the galactic center, after the density-based extraction of the disc stars.}
    \label{fig:compare_1}
\end{subfigure}
\hfill
\begin{subfigure}{0.45\textwidth}
    \centering
    \includegraphics[width=\linewidth]{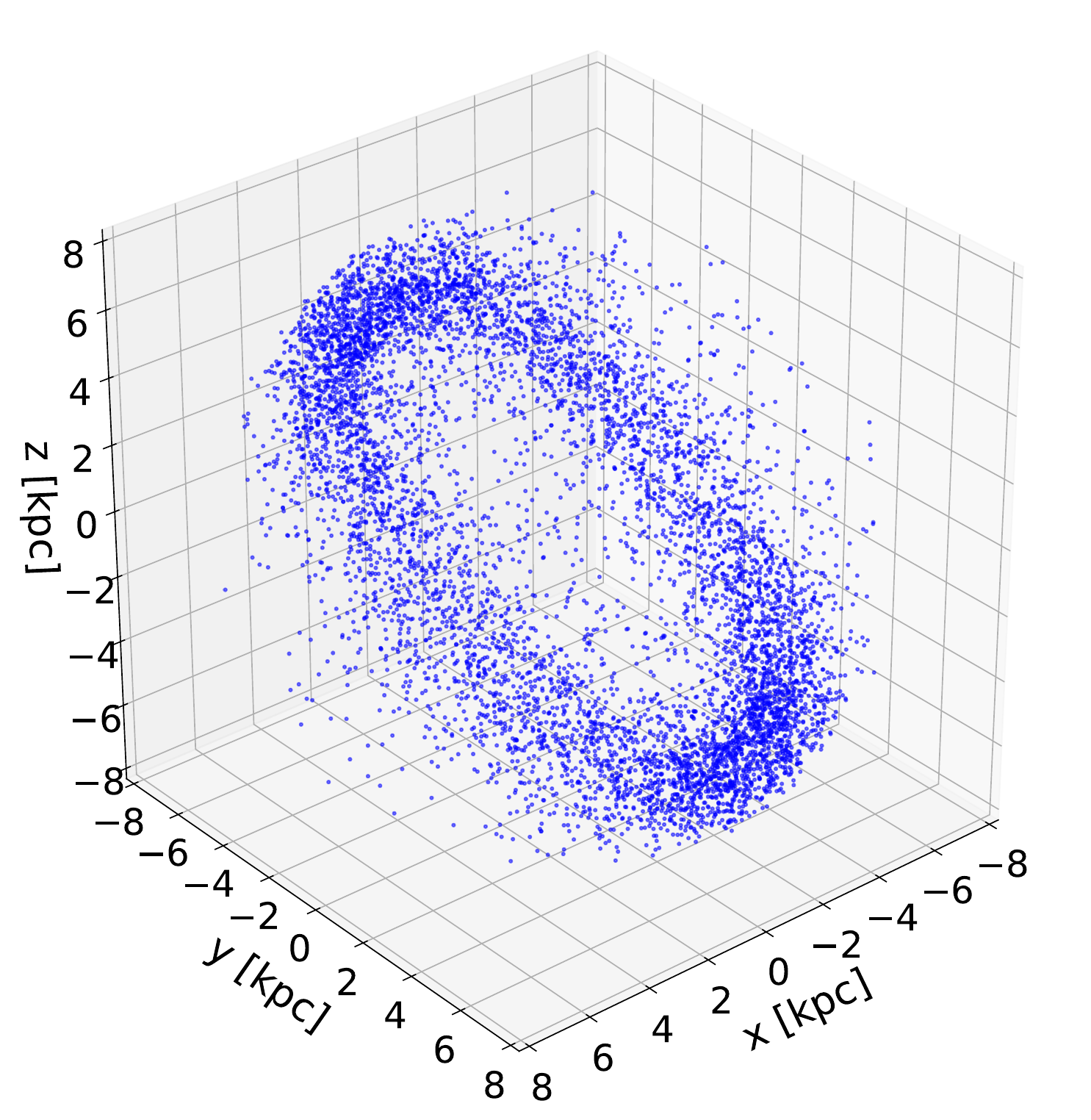}
    \caption{DM particles extracted using stars at $7\text{--}9$ kpc from the galactic center, without the density-based extraction of the disc stars.}
    \label{fig:compare_2}
\end{subfigure}

\caption{Comparison of the three-dimensional distribution of the DM with and without the density-based extraction.}
\label{fig:gaussian_comparison}
\end{figure}

Even if the solar-neighborhood-like region is identified using only the star particles at $7\text{--}9\,\mathrm{kpc}$ from the galactic center, low-density stars belonging to the halo may still be mixed in. We therefore compare the cases with and without the density-based separation of components. Fig.~\ref{fig:gaussian_comparison} shows the three-dimensional distribution of the DM in Method A for these two cases. The density-based separation effectively suppresses the halo contribution and yields a selection that corresponds more closely to the galactic disc. Although its quantitative impact on the local DM density and velocity distribution is limited, we adopt this separation to identify a more realistic solar-neighborhood-like region.

\clearpage

\acknowledgments
We thank Naoki Yoshida for the encouragement at the very early stage of this work.
The work was in part supported by JSPS KAKENHI Grant Numbers No.~24K07061, 24H02244 (KIN), 
and 22K03644 (SM).

\end{document}